\documentclass[twocolumn, nofootinbib,amsfonts,prd,aps]{revtex4}
\usepackage{graphicx}
\usepackage{subfig}
\begin{document}

\title{Quantum shells in a quantum space-time}

\author{Rodolfo Gambini$^{1}$,
Jorge Pullin$^{2}$}
\affiliation {
1. Instituto de F\'{\i}sica, Facultad de Ciencias, 
Igu\'a 4225, esq. Mataojo, 11400 Montevideo, Uruguay. \\
2. Department of Physics and Astronomy, Louisiana State University,
Baton Rouge, LA 70803-4001}

\begin{abstract}
  We study the quantum motion of null shells in the quantum spacetime
  of a black hole in loop quantum gravity. We treat the shells as test
  fields and use an effective dynamics for the propagation
  equations. The shells propagate through the region where the
  singularity was present in the classical black hole space-time, but
  is absent in the quantum space-time, eventually emerging through a
  white hole to a new asymptotic region of the quantum space-time. The
  profiles of the shells get distorted due to the quantum fluctuations
  in the Planckian region that replaces the singularity. The evolution
  of the shells is unitary throughout the whole process.
\end{abstract}

\maketitle
\section{Introduction}
The exact space of solutions of the equations of loop quantum gravity
for vacuum, spherically symmetric space-times was recently found
\cite{spherical}. The breakthrough was due to the realization that a
rescaling and linear combination of the constraints of canonical
quantum gravity in the spherical case yields a constraint algebra that
is a Lie algebra, allowing to complete the Dirac quantization of the
model. The space of physical states can be found in closed form. It is
based on one dimensional spin networks. The proximity of the nodes of
the spin networks is limited by the condition of the quantization of
the areas of the spheres of symmetry. The singularity that is inside
classical black holes is replaced by a region where a description in
terms of a semi-classical geometry is not possible and through it one
propagates into another region of space-time. Here we would like to
analyze the propagation of spherical null shells in the quantum black
hole space-time. We will see that contrary to the behavior of quantum
shells in more traditional treatments \cite{hajicekkiefer}, an ingoing
shell makes it through the quantum region and emerges on the other
side of where the singularity used to be, via a white hole.  Two types
of quantum effects are present when propagating through the region
where the singularity used to be: 1) The relative fluctuations of the
Dirac observable that represents the metric of space-time are maximal;
2) Effects of the quantization of area are also maximal giving rise to
large discontinuities in the value of the Dirac observable that
represents the metric when going from a point of the spin network to
its closest neighbor. This induces random perturbations in the profile
of the shells propagating through that region. We will also show that
shells placed close to the horizon in a quantum state involving the
superposition of ingoing and outgoing shells suffer a phenomenon
analogous to Hawking radiation in that one shell falls into the black
hole and is trapped and the other emerges, due to fluctuations in the
position of the horizon. The evolution of the system is unitary if one
considers both shells but unitarity is lost if one only considers the
outgoing shell. We will treat the shells as test fields to study their
propagation, and then consider their back reaction on the geometry. We
will approximate the evolution through effective semi-classical
equations, to avoid dealing with the true equations of motion that
stem from the quantum space-time, that involve coefficients that
change discretely in value from one link of the spin network to the
next. The approximation is good except in the region where the
singularity used to be where a more careful treatment will be needed
to study the propagation. We only comment on some qualitative features
of the propagation in that region.

\section{Spherical shells on a spherical background}

\subsection{Classical theory}

To treat the spherically symmetric background quantum space time we
choose variables adapted to spherical symmetry, one is left with two
pairs of canonical variables $E^\varphi$, ${K}_\varphi$ and $E^x$,
$K_x$, that are related to the traditional canonical variables in
spherical symmetry $ds^2=\Lambda^2 dx^2+R^2 d\Omega^2$ by
$\Lambda=E^\varphi/\sqrt{|E^x|}$, $P_\Lambda= -\sqrt{|E^x|}K_\varphi$,
$R=\sqrt{|E^x|}$ and $P_R=-2\sqrt{|E^x|} K_x -E^\varphi
K_\varphi/\sqrt{|E^x|}$ where $P_\Lambda, P_R$ are the momenta
canonically conjugate to $\Lambda$ and $R$ respectively, $x$ is the
radial coordinate and $d\Omega^2=d\theta^2+\sin^2\theta
d\varphi^2$. We will take the Immirzi parameter to one.  After doing
the rescaling and combination of the constraints discussed in
\cite{spherical} the total Hamiltonian becomes \cite{hajicekkiefer},
\begin{eqnarray}
H_T &=&\int dx \left[ -N'
  \left(-\sqrt{\vert
      E^x\vert}\left(1+K_\varphi^2\right)+\frac{\left(\left(E^x\right)'\right)^2\sqrt{\vert E^x\vert}}{4
          \left(E^\varphi\right)^2}\right.\right. \nonumber\\
&&+\left. \!\!\!\begin{array}{cc}&\\&\\\end{array} 2 G M\right)+ N
\frac{\left(E^x\right)'\sqrt{\vert E^x\vert} \eta\, p\,
  \delta\left(x-r\right)}{\left(E^\varphi\right)^2}\nonumber\\
&&+2N \frac{K_\varphi \sqrt{\vert E^x\vert}}{E^\varphi}p\,\delta\left(x-r\right)\nonumber\\
&&\left.+ N_r \left[-
(E^x)' K_x +E^\varphi K_\varphi' - p\, \delta\left(x-r\right)\right]\frac{}{}\right] 
\end{eqnarray}
with $r$ the position of the shell and $p$ its canonical momentum. The
parameter $\eta=\pm 1$ is the sign of the momentum, depending on it
one will have shells that are either ingoing or outgoing if one is
outside the black hole. We
will work perturbatively, taking the solution for the space-time
of a black hole without a shell, putting a test shell on it and
studying its back reaction on the geometry. 

\subsection{Quantization}

We proceed to quantize the model. For the gravitational sector we
consider the exact physical states found in \cite{spherical}. On those
states we define an evolving constant of the motion (a Dirac
observable dependent on a (functional) parameters) that represents the matter
Hamiltonian, following a procedure outlined in \cite{hawking}. We
assume the quantum states are a direct product of the gravitational
states and the shell states. We take
the expectation value of the shell Hamiltonian with respect to the
gravitational states and this provides a classical Hamiltonian for
matter that incorporates the corrections of the quantum space-time. We
then proceed to quantize such Hamiltonian using traditional
quantization techniques since it is a mechanical system. We will then
study the back-reaction on the quantum space-time. 

To construct the evolving constant of the motion for the Hamiltonian
we first recall the expression of $\hat{E^x}$ in terms of an evolving
constant of the motion $\hat{E}^x(z(x)) \vert
\tilde{g},\vec{k},M\rangle> =
\hat{O\left(z(x)\right)}\vert \tilde{g},\vec{k},M\rangle> $ with $z(x)$ the
(functional) parameter and the Dirac observable $\hat{O}$ is defined
by $\hat{O}(z) \vert \tilde{g},\vec{k},M\rangle = \ell_{\rm Planck}^2
k_{{\rm Int}\left(z V\right)} \vert \tilde{g},\vec{k},M\rangle$, with
$V$ the number of vertices in the spin network and ${\rm Int}$ means
integer part. In terms of $\hat{E}^x(z(x))$ (a Dirac observable), the
matter part of the Hamiltonian constraint can be written as a Dirac
observable itself,
\begin{equation}\label{4}
  \hat{H}_{\rm shell} =\frac{{\sqrt{\vert \hat{E^x}\vert}}\left(\hat{\left(E^x\right)'}
      \eta +2K_\varphi \hat{E}^\varphi\right)p }{\left(\hat{E}^\varphi\right)^2},
\end{equation}
and $\hat{E^\varphi}$ can be written in terms of $\hat{E}^x$ solving
the Hamiltonian constraint 
\begin{equation}\hat{E}^\varphi=\frac{\left(\hat{E}^x\right)'}{
2\sqrt{1+K_\varphi^2-2GM/\sqrt{\vert \hat{E}^x \vert}}}.
\end{equation}
The
Hamiltonian is an evolving constant of the motion parameterized by the
(functional) parameters $z(x)$ (present in $\hat{E}^x$) and
$K_\varphi$. The first one is associated with spatial diffeomorphisms
and the latter with the slicing of space-time chosen. One can now
evaluate the expectation value of the Hamiltonian of the shell on the
exact physical states of the gravitational theory. At this point it is
good to choose the parameter $K_\varphi$ of the shell Hamiltonian by
fixing $K_\varphi$ to be a (negative) function of $x$ that vanishes
outside the black hole (starting with the first point of the spin
network outside where the classical event horizon would have been) and
is non-vanishing inside, its absolute value taking a maximum value
where the singularity used to be and then diminishing as one goes past
it and eventually becomes zero just outside the Cauchy horizon inside
the black hole. These types of slicings are penetrating slicings akin
to Eddington--Finkelstein coordinates and will allow us to follow the
evolution of shells as they penetrate inside the black hole. 

We will choose the states of the gravitational variables judiciously
so they approximate well a classical space-time in regions far away
from where the classical singularity used to be.  This requires
superpositions of states $\vert \tilde{g},\vec{k},M\rangle$ such that
the values of $\vec{k}$ vary monotonously and without significant
jumps between adjacent values (the values of $\vec{k}$ determine the
radial coordinate up to a Planck scale) and that the spread in the
values of $M$ is small compared to its value. The minimum separation
in radial coordinate in these states is given by $\ell_{\rm
  Planck}^2/(2r)$ with $r$ the radial coordinate such that $4\pi r^2$
is the area of the surfaces of symmetry. In this idealized
context, where one has an eternal black hole and no true dynamics, it
could also be possible to choose a state that approximates a
semi-classical geometry in the region of high curvature. It will have
large discontinuities in the relative values of the metric components
in that region, but well defined values. However, such
a choice appears artificial. Instead we choose states that are
superpositions and given the large (and different) discontinuities in
the region of high curvature for each of the component states one does
not have a semi-classical geometry well approximated there.
Taking the expectation value of the shell Hamiltonian on
the exact physical states of vacuum gravity essentially determines the
prefactor of $p$ in (\ref{4}) entirely as a function $f(r,\eta)$
determined by $z(r)$, $K_\varphi(r)$ and $\eta$.

\section{Motion of the shells on the quantum space-time and backreaction}
From the quantum corrected matter Hamiltonian one can work out the
equations of motion for the shell. For that it will be convenient to
specify the parameters in the parameterized Dirac observables
representing $E^x$ and $E^\varphi$. We have already chosen
$K_\varphi$. We now choose a spin network such that the eigenvalues of
$\hat{E}^x$ are approximated very well by $x^2$, therefore
determining the (functional) parameter $z(x)$.  The condition on $E^x$
implies $N_r=0$ and the diffeomorphism constraint can be solved for
$K_x$. The condition on $K_\varphi$ implies that $N'=0$, so the lapse
is a constant we call $N_0$.  This allows to
solve the Hamiltonian constraint for $E^\varphi$ and therefore the
metric. Taking the Poisson bracket of $r$ and $p$ with the Hamiltonian
we get,
\begin{eqnarray}
  \dot{r}&=& N_0 \frac{\sqrt{\vert E^x(r)\vert }\left(\left(E^x\right)'(r) \eta+ 2
      K_\varphi(r) E^\varphi(r)\right)}{\left(E^\varphi\right)^2},\label{3}\\
  \dot{p}&=& -\frac{\partial \dot{r}}{\partial r} p,\label{4}
\end{eqnarray}
from which it follows that $\dot{r}p$ is a constant and is the
contribution of the shell to the Hamiltonian constraint up to a Dirac
delta function. This allows to combine it with the other terms in the
constraint and what one has is that the $2GM$ term in the Hamiltonian
gets modified by a constant $\dot{r}p$, times a Heaviside function
$H(x-r)$, i.e. the energy of the shell contributes to the mass of the
spacetime outside of the shell. This is the back-reaction of the shell
on the space-time. Notice that the variables $r,p$ are continuous, but
the coefficients of the equations of motion (\ref{3},\ref{4}) will
change discretely as one traverses from one link of the spin network
to the next through the evolution of the shell. In this paper we will
not take this into account and treat the equations as a semi-classical
approximation with the coefficients determined by a continuous metric,
as we are choosing a quantum state in which the jumps in the values of
variables from a link of the spin network to the next are small. We
will discuss separately what happens in the region where the
singularity used to be, where this approximation does not hold.

\subsection{Shells with ingoing momentum}

Depending on the sign of $\eta$ the analysis changes. We start with
$\eta=-1$ which corresponds an infalling shell from outside the black
hole passing through the Planckian region inside and emerging in the
white hole. One can study $f(r,\eta)$ numerically with choices for
$z(r)$ and $K_\varphi$ as the ones we discussed that allow to cover
the entire space-time. It turns out that it can be modeled by a function
$f(r) = -r_0^2/(r^2+\epsilon^2)$ with $r_0$ a macroscopic quantity
much smaller than the mean value of the Schwarzschild radius, since
the exact $f(r)$ is very small at the horizon, and $\epsilon$ is small,
of the order of Planck length. Notice that $r$ now spans from $\infty$
at $i_0$ all the way to $-\infty$ with $r=x=0$ the point where the
singularity used to be in the classical solution. 

We define a self-adjoint operator associated with the
shell's Hamiltonian and determine its spectrum,
\begin{equation}
  -2 i f(r) \Psi(r)'-if(r)'\Psi(r)-2E\Psi(r)=0,
\end{equation}
with $E$ its eigenvalue, which is positive as it is the correction of
the mass due to the presence of the shell.  Its eigenstates can be
found $\Psi_E(r)= C \exp\left(\int dr \frac{-f(r)'+2 i E}{2
    f(r)}\right)$. The integral can be computed in closed form for the
simple form of $f(r)$ we introduced above, $\Psi_E(r)
=\exp\left(-\frac{iE}{r_0^2} \left(r^3/3+\epsilon^2
    r\right)\right)\sqrt{r^2+\epsilon^2}/(\sqrt{2\pi r_0^2 })$. These
states are orthonormal $\int_{-\infty}^\infty dr \Psi_E\Psi_{E'} =
\delta_{E,E'}$ and they satisfy a closure relation, showing that the
Hamiltonian is indeed self-adjoint.  Superposing such states with a
Gaussian weight in $E$, one obtains a time dependent wavefunction for
the propagation of the shell, with a unitary evolution, as the one
shown in fig. 1,
\begin{figure}[h]\label{fig1}
\includegraphics[height=7.5cm]{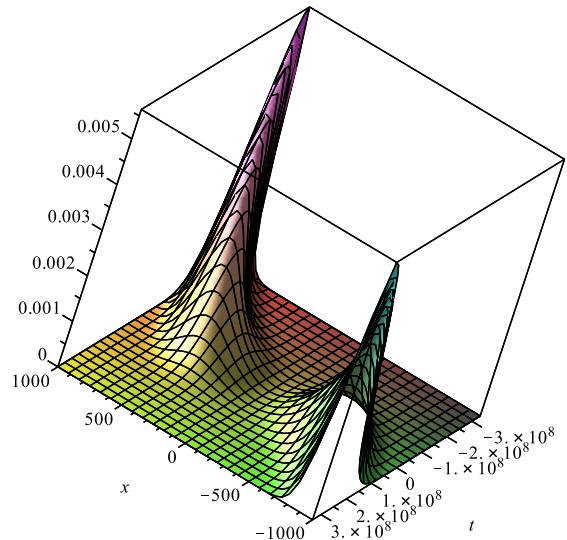}
\caption{The evolution of a shell incoming towards the black hole from
  the exterior ($p<0$), in an approximation where we consider a
  semi-classical geometry everywhere. The amplitude diminishes, but is
  non-vanishing in the Planckian region around $x=0$ for all values of
  $t$. The portion at the left is the shell incoming into the black
  hole and the portion at the right the shell exiting through the
  whitehole. The behavior is completely symmetric around $x=0$ as the
  effective model considered does not incorporate the fluctuations at
  the Planckian region.}
\end{figure}
Notice that the incoming shell traverses the Planck region and will eventually 
emerge into the new asymptotically flat region that the eternal
non-singular black hole solution in loop quantum gravity has after
traversing a white hole. The eternal non-singular solution is
symmetric around $x=0$ and the horizon of the black hole is mirrored
in $x<0$ by a white hole. As we will see, when one considers a more
realistic evolution through the Planck region, although the space-time
is symmetric around $x=0$ the evolution of the pulse will not be due
to quantum fluctuations.

\subsection{Shells with outgoing momentum}
We also wish to consider the case of a shell with outgoing momentum
($p>0$).  The resulting shell would be outgoing in its motion
($\dot{r}>0$) if it is in the black hole exterior, but would still be
ingoing in its motion, as expected, in the black hole interior.  In
this case we do not have a simple expression approximating the
behavior of $f(r)$ as before, but its behavior can be studied
numerically. One now has that $f(r)$ vanishes at the horizon. That
means that the zero of energies for the shells is placed at the
horizon. Shells with outgoing momentum inside the horizon will have
negative energy and shells outside have positive energies
\cite{draythooft}. This can be easily seen in eq. (\ref{4}):
$(E^x)'>0$ everywhere, in this case $\eta=1$ and $K_\varphi=0$ outside
so one has the first term multiplied times $p>0$ and the energy is
positive.  At the horizon one can check using the definition of
$E^\varphi$ that the two terms in (\ref{4}) cancel each other. From
there inward, their difference is negative as the modulus of
$K_\varphi$ grows. Shells with outgoing momentum inside the horizon
behave like positive energy shells moving into the past.
\begin{figure}[h]
\includegraphics[height=3.5cm]{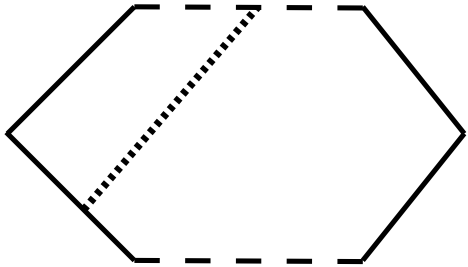}\label{fig1}
\caption{A shell with outgoing momentum inside the horizon has
  negative energy and behaves like a positive energy shell moving into
the past. In our quantum space-time the shell will continue into the
future beyond the Planckian region.}
\end{figure}
Notice that the foliation we are using, which behaves like
Schwarzschild coordinates outside the black hole with $K_\varphi=0$
and behaves like Eddington--Finkelstein coordinates in the interior
labels the hypersurfaces with the time of an observer at infinity. As
such, the foliation cannot go beyond the Cauchy horizon in the
interior of the black hole and the evolution of the shells seems to
``stop'' there. 

\begin{figure}[h]
  \centering
    \subfloat[Initial pulse]{{\includegraphics[height=5.5cm]{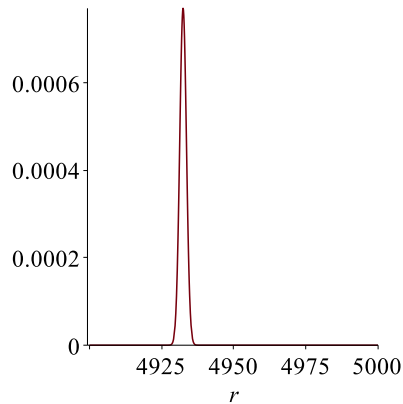}}}
    \qquad
    \subfloat[Final pulse]{{\includegraphics[height=5.5cm]{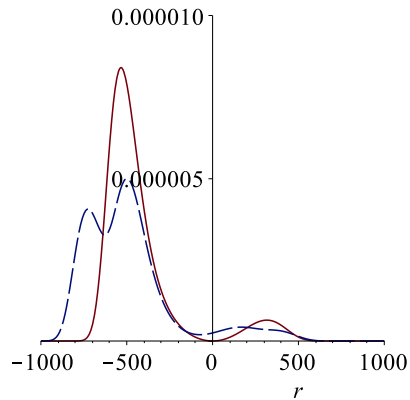} }}
    \label{fig3}

\label{fig3}
\caption{To try to mimic the effect of the fluctuations in the
  Planckian region on the shell we studied the evolution of the same
  shell in two quantum geometries that could be involved in a
  superposed state of the gravitational field and then superposed the
  results for the profiles of the shells. Panel (a) is the incoming
  pulse and in panel (b) we depict together the pulse evolved through
  the Planck regime using a single semi-classical geometry and using
  the superposition of the evolutions in two semi-classical
  geometries. The latter is an attempt to mimic the quantum
  fluctuations. As we see, the pulse is significantly distorted in its
  passage through the Planck region. Notice the significant decrease
  in amplitude in both cases, due to the pulse becoming very spread
  out while passing through the Planck region due to tidal forces. In
  addition to that, the fluctuations produce lasting distortions in
  the pulse after passage. The horizontal axis in in centimeters and
  the mass scale is that of a solar sized black hole. }
\end{figure}

\subsection{Effects of fluctuations in the Planck regime}
Up to now we have made a treatment in terms of a semi-classical metric
even in the Planckian region. Although states that approximate a
semi-classical geometry everywhere do exist in our highly idealized
description of an ever existing black hole, they are likely not the
ones that occur in nature after gravitational collapse. More likely is
that one will have superpositions of gravitational states and that
implies that in the Planckian regions there will be large
fluctuations, for instance, in the curvature. To try to mimic the
effect of these fluctuations in the evolution of shells, we have
studied the evolution of the same shell in two
different quantum states of the gravitational that could enter in such
superpositions and then superposed the evolutions. The results are
shown in figure 3.

\section{Discussion} 

One has interesting consequences when one considers superpositions of
shells near the horizon. Consider outgoing shells. The ones that lie
outside the horizon will make it to scri+. The ones inside the
horizon, in spite of being ``outgoing'' end up traveling through the
Planckian region inside the black hole. Now, in a quantum space-time
one expects fluctuations in the position of the horizon. The horizon
carries out a random walk. After some time the fluctuations
accumulate, and by Page time could imply deviations in the radial
coordinate position of the horizon of the order of Planck length
\cite{whitehole}. A related discussion of fluctuations can be found in
\cite{hu}. That sounds small, but due to the high blueshifts at the
horizons, it implies very different evolutions for shells outside the
horizon depending on where they lie with respect to the
horizon. Therefore if one considers a shell of finite width obtained
by superposing shells like the ones we consider in this paper, the
various components will evolve in very different ways. This suggests a
mechanism for the quantum space-time to produce appreciable effects in
a region where one would have ordinarily not expected significant
corrections due to quantum gravity. Note that although one would have
expected a semi-classical metric at the horizon would describe well 
the situation, given that it is a region of low curvature, it
is the fluctuations that give rise to the effect. This could be a
potential source of breakdown of complementarity.

The fluctuations of the horizon might also imply that some of the shells
in a superposition end up being inside the horizon and will therefore
end up in the Planckian region even if they were outgoing shells
($p>0$).  This is similar to what happens in the pair production
process of Hawking radiation. Notice that the shell inside the horizon
contributes negatively to the mass of the black hole, as is commonly
argued in the heuristic picture of Hawking radiation. 
The evolution of the complete wavepacket of
shells is unitary provided one keeps track of the shells inside the
black hole. Looking at the evolution of the outside ones only,
unitarity would appear lost.

It is worthwhile comparing our results with those obtained with more
traditional quantizations of shells. Most efforts have concentrated on
self-gravitating shells so the parallels are not easy to
draw. However, in those proposals shells ``bounce'' and behaviors like
a superposition of white and black holes emerged leading to the
concept of ``grey holes'' \cite{hajicekkiefer,hajicekkiefer2}. None of
these behaviors are observed here, largely in part due to the
elimination of the singularity in loop quantum gravity. There have
been studies of shells propagating on black holes but have not
addressed in detail what happens at the singularity
\cite{krausswilczek}. The fact that the shells emerge through a white
hole may suggest that this is a way in which information could leak
from an evaporating black hole, although our eternal black hole exact
solution does not allow us to model that situation in detail. A
conjectural scenario based on this observation is presented in
\cite{whitehole}.

\section*{Acknowledgement}
This work was supported in part by Grant
No. NSF-PHY-1305000, funds of the Hearne Institute for Theoretical
Physics, CCT-LSU, and Pedeciba.

\end{document}